# Macrodipoles of potassium and chloride ion channels as revealed by electronic structure calculations

**Fabio Pichierri**

G-COE Laboratory, Department of Applied Chemistry,

Graduate School of Engineering, Tohoku University,

Aoba-yama 6-6-07, Sendai 980-8579, Japan

E-mail: fabio@che.tohoku.ac.jp

**Abstract**

With the aid of quantum mechanical calculations we investigate the electronic structure of the full length (FL) potassium channel protein, FL-KcsA, in its closed conformation, and the electronic structure of the ClC chloride channel. The results indicate that both ion channels are strongly polarized towards the extracellular region with respect to the membrane mean plane. FL-KcsA possesses an electric dipole moment of magnitude 403 Debye while ClC has a macrodipole whose magnitude is about five times larger, 1983 Debye, thereby contributing to differentiate their membrane electric barriers. The dipole vectors of both proteins are aligned along the corresponding selectivity filters. This result suggests that potassium and chloride ion channels are not passive with respect to the movement of ions across the membrane and the ionic motion might be partially driven by the electric field of the protein in conjunction with the electrochemical potential of the membrane.

## 1. Introduction

Biological cells are delimited by phospholipid membranes which embed membrane proteins and other molecules [1]. Many of these membrane proteins are involved in the transmission of signals (cell signaling) or the transport of ions and small molecules from the outside to the inside of cells. Ion channels are proteins specialized in the transport of positively and negatively charged ions across the membrane [2]. Mutations that affect the functionality of ion channels generate a number of diseases termed channelopaties [3,4].

Biophysical studies on ion channels can be traced back to the early 1950s, when Hodgkin and Huxley [5] conducted experimental work on the giant axon of Loligo. Further progress in the field of membrane biophysics was achieved by Neher and Sakmann [6] with the invention of the patch clamp technique which allows the measurement of electric currents associated to ionic fluxes in single ion channels. However, up until the end of the 20th century structural information was missing with the result that a molecular level picture of the ion selectivity and gating (i.e. the opening and closing of channels) processes could not be discussed in detail. It was only in 1998 that the crystal structure of the KcsA was revealed from the first time by MacKinnon and coworkers [7] who succeeded in the difficult task of crystallizing these important proteins. Hence, this study made possible to address for the first time the molecular mechanism of $K^+$ conduction and selectivity by the cell membrane. This breakthrough was soon followed (in 2002) by the structural determination of the ClC chloride channel, a membrane protein which is specialized in the transport of $Cl^-$ ions across the membrane [8].

The crystallographic analyses of both K and Cl channels sparked a new wave of molecular-level studies such as the crystallization of KcsA channels containing different types of cations blocked by tetrabutylammonium [9] and complexes with channel blockers like the peptide Charibdotoxin [10]. The newest achievement in this field is represented by the structural determination of the full length (FL) KcsA channel made by Perozo, Kossiakoff and their coworkers [11], which provides further atomic-level details on this ion channel with respect to the truncated forms. In addition, molecular dynamics (MD) simulations have shed light on the conformational transitions associated to the gating mechanism of several ion channels [12].

Here we investigate the electronic structure of the FL KcsA channel structure and compare it against that of the ClC channel. The main goal of this paper is quantify the degree of electronic charge polarization of these membrane proteins which may help to increase our knowledge about the biophysics of ion channels. Useful information about the polarization of electronic charge can be obtained from the macrodipole to calculate which a quantum mechanical (QM) treatment of the whole protein is necessary [13]. Such calculations are nowadays feasible thanks to the development of novel and efficient algorithms for the quantum mechanical study of molecules containing several thousand atoms [14]. For instance, by using these novel computational methods the author has already investigated the electronic structure of α-chymotrypsin [15], the SH2 domain of p56$^{lck}$ tyrosine kinase [16], ubiquitin [17], and human erythropoietin [18]. In particular, the study of α-chymotrypsin was undertaken in order to assess the quality of the computed macrodipole against experiments which were reasonably well reproduced by our QM calculations [15].

As far as ion channels are concerned, few QM studies have been performed so far. Blyzniuk and coworkers [19,20] performed QM studies of TF-KcsA (PDB ID: 1BL8) with the aim of assessing its electrostatic potential of the selectivity filter. More recently, Stewart has performed a QM calculation of TF-KcsA (PDB ID: 1JVM) [21] and successfully reproduced the coordination to $K^+$ in the selectivity filter. Theoretical studies where models comprising a selected number of amino acid residues were considered have been also carried out for KcsA [22,23].

## 2. Molecular models and methods

The atomic coordinates of the membrane proteins investigated herein were downloaded from the Protein Databank (PDB) [24]. The crystal structure of the FL-KcsA-Fab2 complex (PDB ID: 3EFF) [11] was modified as follows. The Fab2 units (chains A, B, C, D) were removed and hydrogen atoms were added to saturate the open valences of the C, N, O, and S atoms of chains K, L, M, and N in the crystal structure. The terminals of peptide chains were transformed in the zwitterion form. The resulting atomic model of FL-KcsA was subjected to 100 cycles of molecular mechanics (MM) minimization with the Amber 99 force-field as implemented in the Hyperchem software package [25]. This structural relaxation was performed so as to remove bond strains and steric clashes among closely placed amino acid residues while keeping the whole secondary, tertiary, and quaternary structure of the protein as close as possible to the original experimental structure.

The crystal structure of ClC Chloride channel from *E.coli* (PDB ID: 1OTS) [26] was selected and modified as follows. The Fab fragments (chains C, E, D, F) were removed as well as the four $Cl^-$ ions located in the selectivity filter and water molecules. Hydrogen atoms were added to the C, N, O, and S atoms of chains A and B and the C- and N-terminals of the peptide chains were transformed in the zwitterion form. The resulting atomic model of FL-KcsA was then subjected to 100 cycles of MM minimization as done for FL-KcsA.

Electronic structure calculations were performed with the MOPAC2009 software package of Stewart [27]. This software implements the Mozyme algorithm which allows the quantum mechanical treatment of molecular systems containing several thousand atoms through the use of localized molecular orbitals (LMOs) [14]. The new PM6 parameterization was employed here [28].

## 3. Results and discussion

### *3.1 Full-length KcsA potassium channel*

The atomic model of FL-KcsA possesses 8796 atoms and 552 peptide linkages. The Lewis structure of the system is made of 8896 σ-type bonds, 2340 lone-pairs, 728 normal π-type bonds, 96 aromatic ring π-type bonds, 84 positive charges and 56 negative charges and the resulting total charge is +28. As far as the charged residues are concerned, each of the four chains possesses 20 positively-charged amino acid residues comprising 15 arginines, 4 histidines, and one lysine all of which contribute to the large positive charge of the protein. The negatively charged residues in each chain are in total 13, namely 9 glutamate groups and 4 aspartate groups. In addition, each chain possesses a positively charged N-terminal (with Ser22 the first amino acid in the chain) and a negatively charged C-terminal (with Arg160 the last residue in the chain).

The distribution of charged residues within the 3D structure of FL-KcsA is shown in Figure 1. It is interesting to note that the cavity below the selectivity filter does not possess charged residues. Mackinnon suggested that this cavity is useful to overcome the dielectric barrier that an ion experiences when crossing the membrane [7,8].

Furthermore, we notice that the long α-helices in the inner part of the cell are rich in positively-charged arginine residues which contribute to the overall electric polarization of FL-KcsA.

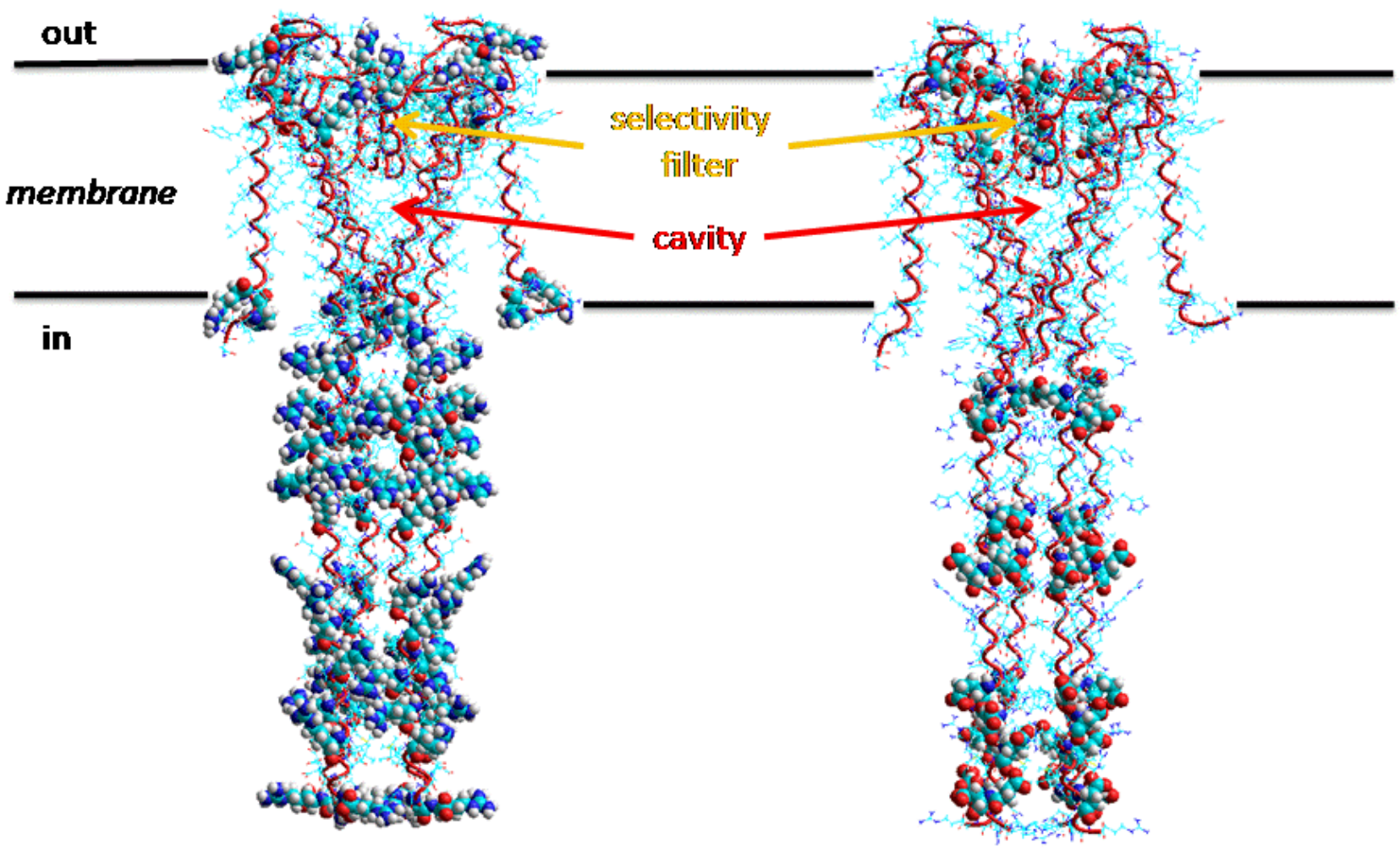

Figure 1. Distribution of positive (left) and negative (right) amino acids in FL-Kcsa.

Figure 2 shows the computed electric dipole moment vector of FL-KcsA which is represented by a yellow arrow. The center of the arrow is placed on the centre of mass of the protein; the arrow's tip indicates the positively-charged end whereas the base of the arrow represents the negatively-charged end. As seen from the figure, the dipole vector is almost normal to the membrane mean plane and aligned along the main pseudo fourfold axis of the protein. The magnitude of the electric dipole moment is 403 Debye, which is somewhat smaller than that of ~ 500 Debye computed early by the author for α-chymotrypsin [15] (note, however, that the present calculation on KcsA does not include the effect of the solvent field on the protein). To the best of our knowledge, neither an experimental nor a theoretical value of the dipole moment of FL-KcsA is as yet available for comparison. Our result may have interesting implications for the biological function of KcsA. In this regard, it highly likely that the macrodipole of KcsA modifies the membrane dielectric barrier thereby allowing $K^+$ ions to cross it.

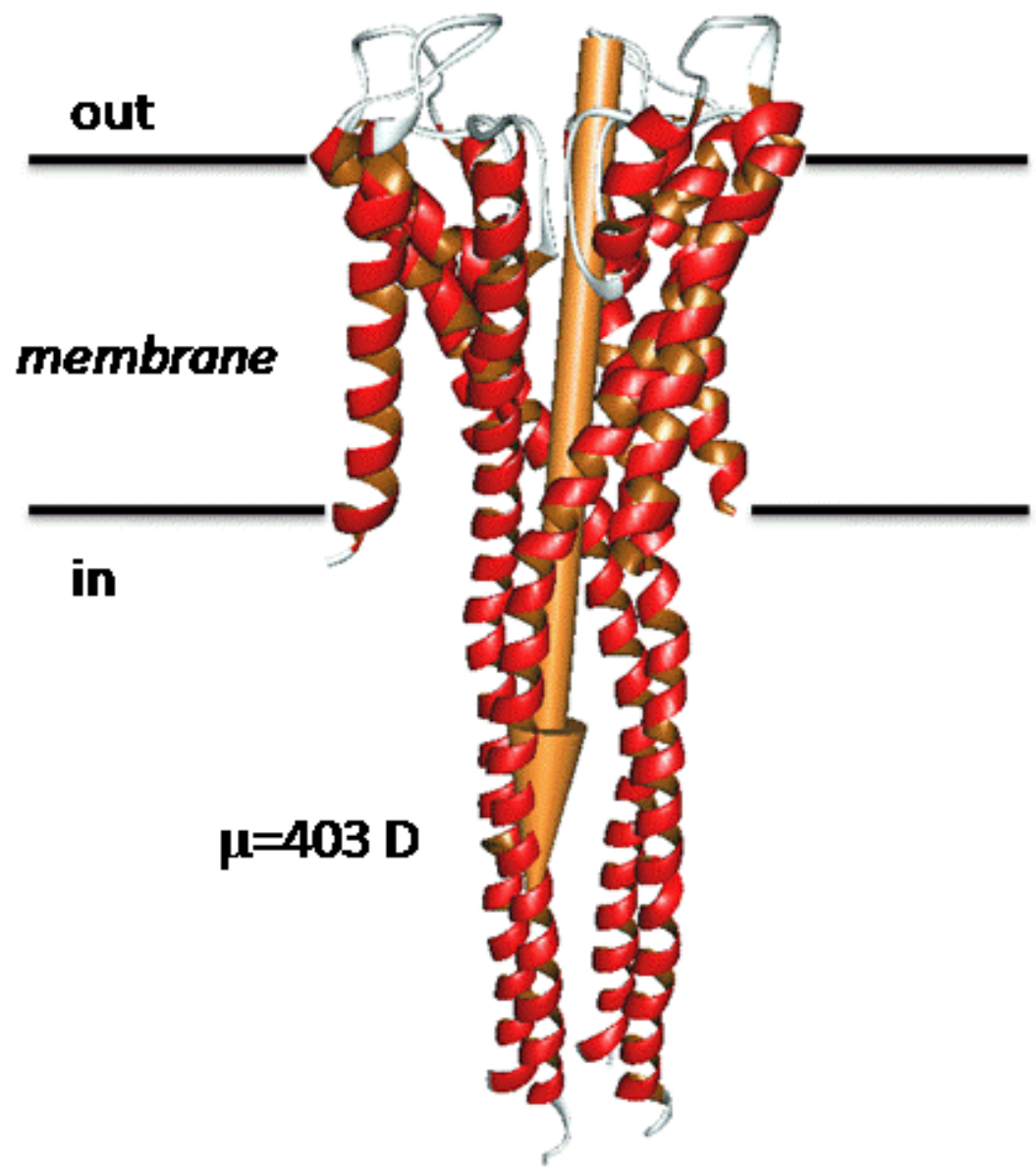

Figure 2. Macrodipole of FL-KcsA.

*3.2 ClC chloride channel*

The atomic model of ClC possesses 13598 atoms and 892 peptide linkages. The Lewis structure of the system is made of 13742 σ-type bonds, 3419 lone-pairs, 1149 normal π-type bonds, 108 aromatic π-type bonds, 77 positive charges and 51 negative charges and the resulting total charge is +26. As far as charged residues are concerned, most of the positive charge arises from arginine residues which are concentrated at the inner membrane surface, in contact with the cytoplasm, as shown in Figure 3. Negatively-charged residues, on the other hand, are roughly equally distributed near both membrane surfaces, above and below the selectivity filter.

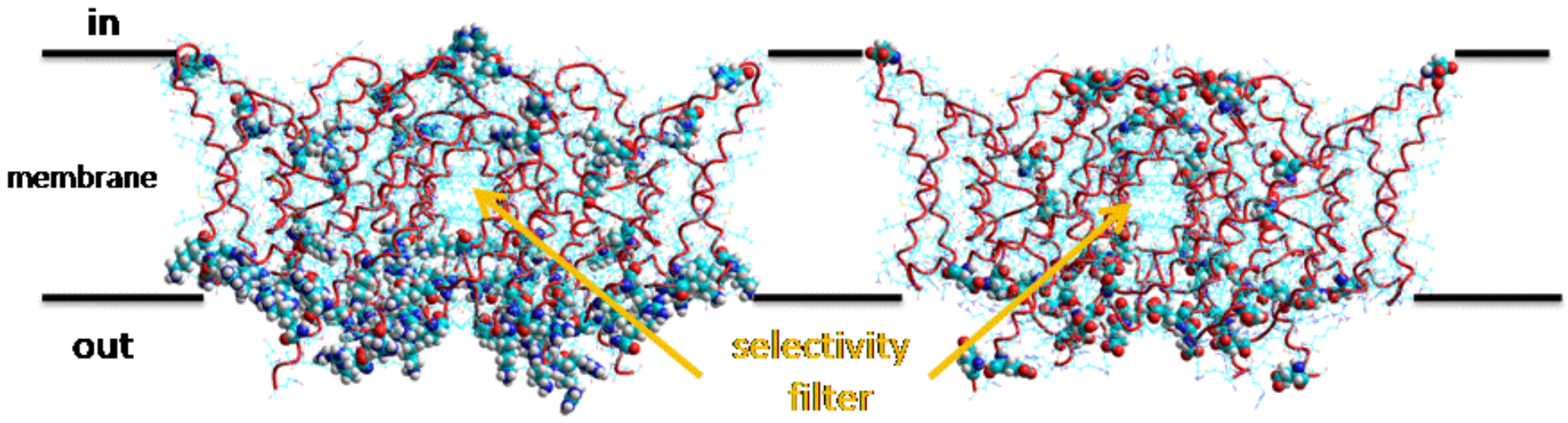

Figure 3. Distribution of positive (left) and negative (right) amino acids in ClC.

Figure 4 shows the electric dipole moment vector of ClC. As for KcsA, the center of the arrow is placed on the centre of mass of the protein while the arrow's base and tip indicate the negative and positive side of the dipole moment vector, respectively. Interestingly, as observed for KcsA, also the dipole vector of ClC is almost perpendicular with respect to the membrane mean plane and with the positive side oriented toward the internal part of the cell. The magnitude of the macrodipole of ClC, however, is about five times larger than that of KcsA, being 1983 Debye. Macrodipoles of such magnitude are not uncommon and, as noticed by Porschke, the dipole vector of membrane proteins may sustain vectorial ion transport across the membrane [29]. Our result supports this view and, in addition, it indicates that the membrane dielectric barrier of ClC is different from that of KcsA. Therefore, it appears that a strongly polarized ion channel may be necessary in order to overcome the electrochemical potential and hence allow the transfer of anions from the outside to the inside of the cell.

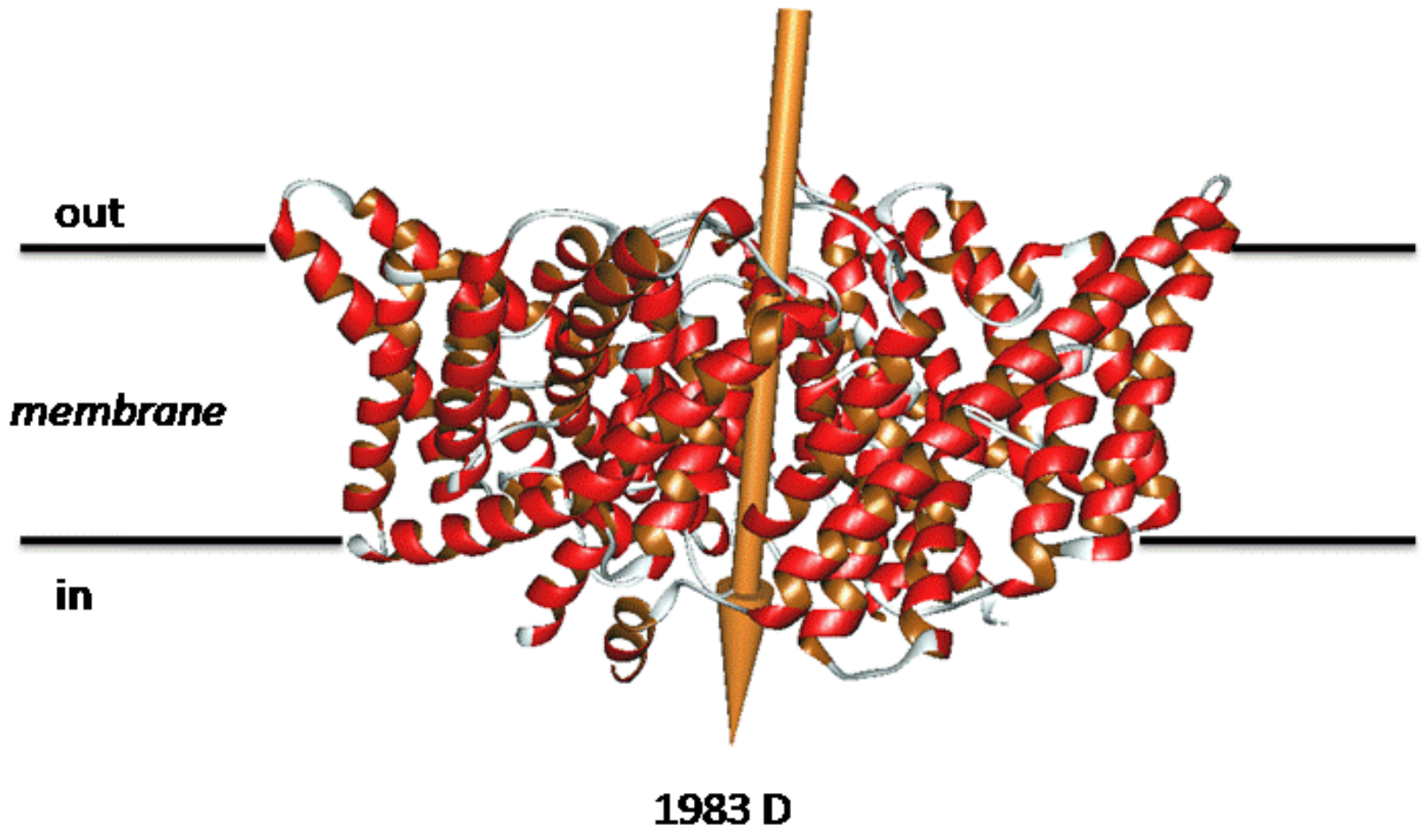

Figure 4. Macrodipole of ClC chloride channel.

## 4. Conclusions

We carried out a quantum mechanical study of full-length KcsA potassium and ClC chloride channels and compared their electronic structures. Both proteins possess a macrodipole which is oriented along their respective selectivity filters and indicating that their electronic charge is strongly polarized towards the outer of the cell. The magnitude of the macrodipoles, however, is significantly different with the ClC macrodipole (1983 D) being almost five times larger than the FL-KcsA macrodipole (403 D). Hence, it appears that the membrane dielectric barriers of these proteins are significantly different.

## Acknowledgments

I thank Dr. James JP Stewart for a copy of MOPAC2009 and for many useful advices on the proper usage of the software. The Global COE program (IREMC) and the Graduate School of Engineering of Tohoku University are gratefully acknowledged for financial support. This work was supported by Grant-in-Aid for Scientific Research [Kakenhi C 20550146].